# Charge transfer at hybrid inorganic-organic interfaces

Xiaoming Wang[1], Keivan Esfarjani [2,3] and Mona Zebarjadi[1, 2, 4]

[1]Institute for Advanced Materials, Devices and Nanotechnology, [2]Department of Mechanical and Aerospace Engineering, Rutgers University, Piscataway, New Jersey 08854, United States

[3]Department of Mechanical and Aerospace Engineering, [4]Electrical and Computer Engineering Department, University of Virginia, Charlottesville, Virginia 22904, United States

e-mail address: m.zebarjadi@virginia.edu

Organic dopants are frequently used to surface-dope inorganic semiconductors. The resulted hybrid inorganic-organic materials have a crucial role in advanced functional materials and semiconductor devices. In this article, we study charge transfer at hybrid silicon-molecule interfaces theoretically. The idea is to filter out the best molecular acceptors to dope silicon with hole densities as high as $10^{13}$ cm$^{-2}$. Here, we use a combinatorial algorithm merging chemical hardness method and first-principles DFT and GW calculations. We start by using the chemical hardness method which is simple and fast to narrow down our search for molecular dopants. Then, for the most optimistic candidates, we perform first-principles DFT calculations and discuss the necessity of GW corrections. This screening approach is quite general and applicable to other hybrid interfaces.

## I. INTRODUCTION

Hybrid inorganic-organic (HIO) materials are attracting in many fields such as field-effect transistors (FETs) [1-3], solar cells [4-6] and thermoelectrics [7-9] owing to their low cost, easy processing and material abundance. Self-assembled monolayer (SAM) of molecules are usually used in FETs as high-capacitance dielectrics due to their insulating properties.[1,2] The molecular SAMs are also found to be potential dopants to tune the carrier densities of semiconductor devices.[10-13] Recently, HIO materials are introduced in the thermoelectric field. For example, by intercalation of molecules, layered transition metal dichalcogenide TiS$_2$ demonstrate promising ZT of 0.28 at 373 K due to enhancement of the power factor and reduction of the thermal conductivity.[9]

Silicon is advantageous as the inorganic part in HIO devices due to its high performance in integrated circuit electronics and mature fabrication technology. Within the current photovoltaic technology, silicon based solar cells dominate the market with over 80% share.[14] In addition, nanostructured silicon is promising for efficient thermoelectrics. Bulk silicon has large power factor of ~4 mW/(mK$^2$)[15] at room temperature. Using nano-structuring[16-18] or adding holes[19], the thermal conductivity could be reduced to as low as 2-3 W/(mK), which allows silicon to have a ZT of 0.4 at room temperatures and ZT values as high as 0.95 at 900 C$^0$.[16,19]

The basic operation principles of HIO devices are based on charge transfer between an inorganic body and organic compound at the interface. After contact, chemical potentials should align resulting in bending of the bands and charge accumulation near the interface and surface doping of the inorganic material[11,12,20]. Charges are confined normal to the interface but free to move in the parallel direction. One can expand such 2D geometry to a 3D geometry and use nanowire forest surface doped with molecules or periodic holely structures with surface doped holes. In that case if the distance between the dopants is less than the screening length, then it is possible to have free carriers which are not tightly bound to the surface interface, and could travel within the bulk of the matrix to lower surface scattering. The high mobility as a result of surface transfer doping, free of ionized impurity scattering, could improve the performance of the electronic devices and was also found able to enhance the performance of bulk thermoelectric materials.[21,22]

It is crucial to maximize the charge transfer in order to achieve good performance of the HIO devices. The propensity of charge transfer at the interface requires proper band alignments of the participating organic and inorganic materials. Take p-doped silicon with molecular dopants as an example. The charge transfer barrier for hole doping of Si, $\Delta$ is defined as $\Delta = EA - IP$, where EA is the electron affinity of the molecule and IP is the ionization potential of silicon. Charge transfer occurs spontaneously in the ground state whenever $\Delta$ is negative, whereas positive values of $\Delta$ indicate charge transfer excitations. To optimize the charge transfer of silicon-molecule interfaces, one needs to screen the molecules to find the smallest $\Delta$. In order to apply surface transfer doping, the silicon surface should be passivated to prevent charge confinement by the surface states. Previous study has seen efficient n-type doping of silicon by cobaltocene with transferred electron density of $2 \times 10^{13}$ cm$^{-2}$.[13] Several other studies[23-27] show that strong acceptors such as tetracyanoquinodimethane (TCNQ) and 2,3,5,6-tetrafluoro-7,7,8,8-tetracyanoquinodimethane (F4TCNQ) could withdraw electrons from silicon resulting in p-type doping. However, questions like how many charges per molecule are transferred in these HIO systems or which molecule is the best to efficiency inject holes to silicon, are not addressed. Herein, we systematically study the charge transfer between silicon and a diversity of molecules by using a combinatorial algorithm merging the chemical hardness method and the first-principles density functional theory (DFT) and many-body Green's function GW formalism. For thermoelectric devices both n-doped and p-doped legs are needed and indeed both types of HIO materials exist. Whereas for p-type doing there are many dopant candidates, for n-type doping there is only one in literature which results in surprisingly high doping concentration[13]. Thus we focus only on p-type doping in this article. We screen out the best candidates for charge transfer using the simple and approximate models, and then perform state-of-the-art first-principles DFT and many-body GW calculations to accurately address the band alignments and charge transfer of the screened candidates. We find that both F4TCNQ and 3,5-difluoro-2,5,7,7,8,8-hexacyanoquinodimethane (F2HCNQ) with strong electron-withdrawing ability can effectively dope silicon with hole densities as high as $10^{13}$ cm$^{-2}$.

## II. METHODOLOGY

A quick and approximate scan of different molecules for charge transfer in silicon-organic systems is performed by using the chemical hardness method. Then, through first-principles DFT and GW calculations, we provide a more accurate description of the band alignment and charge transfer in the filtered top candidates.

## Chemical hardness method

For silicon-molecule HIO materials, we employ a thermodynamic model called the chemical hardness method, to predict the charge transfer. In this model, the total energy of a system is expanded in powers of the net charge ($\delta n$) as follows:

$$E(\delta n) = E_0 + \frac{\partial E}{\partial (\delta n)} \delta n + \frac{1}{2} \frac{\partial^2 E}{\partial (\delta n)^2} (\delta n)^2 + \cdots$$
$$= E_0 + \mu \delta n + \frac{1}{2} \eta (\delta n)^2 + \cdots \quad (1)$$

Where $E_0$ is the total energy of the neutral system. By definition, the first derivative of the total energy is the chemical potential ($\mu$) and the second derivative is called the chemical hardness ($\eta$)[28]. As its name indicates, materials with large chemical hardness do not easily allow charge transfer when put in contact with other materials. We assume that after contact, the chemical hardness of the two materials are, to leading order, not modified due to deformations or charge transfer. Clearly, in the case of chemisorption, where hybridization takes place, this is not a good approximation. But in the physisorbed case, such assumption is realistic. Note that in the case of chemisorption, the question of how much charge is transferred is not a well-defined question as there are shared electrons at the surface bonds. Therefore, this study is only focused on the physisorbed cases.

Assume after contact, charge $\delta n$ is transferred from B to A as a result of the chemical potential difference between A and B. Then the total energy of the A/B system, could be written as the sum of the total energies of the isolated charged systems (A and B) and the energy of the resulted charged capacitor formed between A and B:

$$E = E_A + E_B + (\delta n)^2 / 2 C_{AB}$$
$$= E_{0,A} - \mu_A \delta n + \frac{1}{2} \eta_A (-\delta n)^2 \quad (2)$$
$$+ E_{0,B} + \mu_B \delta n + \frac{1}{2} \eta_B (\delta n)^2 + (\delta n)^2 / 2 C_{AB}$$

After the two systems are put in contact with each other, the charge transfer is determined by minimization of the total energy with respect to $\delta n$.

$$\delta n = \frac{\mu_A - \mu_B}{\eta_A + \eta_B + 1/C_{AB}} \quad (3)$$

where the coefficient $C_{AB}$ is the mutual capacitance of the two subsystems, which can be approximately evaluated using the following formula:

$$1/C_{AB} = e^2 \sum_{i \in A, j \in B} \frac{|c_i|^2 |c_j|^2}{4\pi\varepsilon |\mathbf{r}_i - \mathbf{r}_j|} \quad (4)$$

Where $\varepsilon$ is the vacuum permittivity, $\mathbf{r}_i$ and $\mathbf{r}_j$ are the positions of atoms belonging to the A and B subsystems, respectively. In the linear combination of atomic orbitals (LCAO) framework, $c_i$ is the coefficient of atomic orbital ($\phi_i$) centered on atom $i$:

$$\psi_{HOMO/LUMO} = \sum_{i=1}^{n} c_i \phi_i$$
$$\sum_{i=1}^{n} |c_i|^2 = 1$$

(5)

The chemical hardness method was once used to calculate the charge transfer in hetero-atomic molecules and hetero-molecular dimers, but extended for solid interfaces later[29,30]. In solid materials, when a gap exists, the chemical hardness is equivalent to the bandgap. The capacitance term $1/C_{AB}$ in Eq. (3) is usually small compared to typical band gaps and can be calculated as the Coulomb interaction between the LUMO of the acceptor and the HOMO of the donor.

## First-principles DFT and GW calculations

Though the chemical hardness method could give a qualitative and intuitive estimation of the charge transfer within several simplifications and approximations, the main drawback of the method is neglecting the physics and chemistry at real interfaces. DFT based first-principles calculations can take into account the interface structure relaxations which is believed to be important for charge transfer. However, the dynamic polarization effects which have a great impact on the electronic levels of molecules and semiconductors are not captured by normal DFT or even hybrid functional calculations.[31,32] To this end, we resort to many-body perturbation theory (MBPT) in its GW approximation[33-35] for more accurate description of the band alignments and charge transfer at silicon-organic interfaces. MBPT provides a systematic method to obtain the true single-particle excitations from the Green's function of the system. All the GW calculations in the present study are single-shot $G_0W_0$ calculations which use normal DFT (PBE in this work) single-particle states as input. The wave functions are not updated during the GW calculations and only the orbital energies are corrected. We perform DFT and GW calculations on F2HCNQ-, F4TCNQ- and TCNQ-Silicon HIO systems. The slab model is adopted to represent the interface structure.

To perform the DFT calculations, we employ the Quantum ESPRESSO package[36]. The PBE[37] based generalized gradient approximation (GGA) functional is used for the exchange-correlation energy. The ion-electron interactions are treated using the ONCV[38,39] norm-conserving pseudopotentials. The cutoff energy of the plane wave basis set is 60 Ry. Monkhorst-Pack[40] k meshes of 3×3×1 and 8×4×1 are used to sample the Brillouin zone for the unit cell of the parallel orientation (PO) and nitrogen-terminated orientation (NO) configurations, respectively, see Fig. 1. A vacuum thickness of 12 Å is used to eliminate the periodic image interactions. To correctly deal with the van der Waals interactions, we employ the non-local DFT functional vdW-DF-cx[41] to relax the structure. The top silicon layers, the interface hydrogen atoms and the molecules are allowed to relax while the bottom silicon and hydrogen atoms in the silicon slab are fixed to the pre-relaxed positions. The force convergence criteria is 1e-4 Ry/Bohr$^2$.

The GW calculations are performed only in Γ-point using the WEST code[42], based on DFT optimized structure. In the spectral decomposition of the dielectric matrix we include $N_{PDEP}$=1024 eigenvectors. Supercells of 11.541×11.541×21.88 Å$^3$ and 15.388×15.388×28.16 Å$^3$ are constructed for the PO and NO configurations, respectively, see Fig. 1.

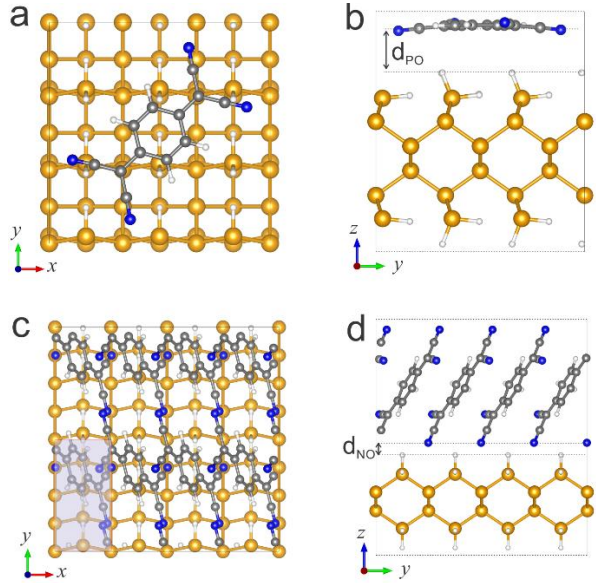

FIG. 1 The schematic configurations of the Si(100):H-TCNQ interface. (a) Top and (b) side view of the parallel orientation (PO). (c) Top and (d) side view of the nitrogen-terminated orientation (NO). For PO, we perform both PBE and GW calculations on the unit cell shown in (a), while (c) shows the supercell in GW calculation for NO, the shaded area displays the unit cell in PBE calculation. The yellow, grey, white and blue balls denote the Si, C, H and N atoms, respectively.

In the slab model, the slab thickness should be large enough to converge the properties to the bulk values. To this end, we investigate the charge transfer barrier variation of Si-TCNQ interface by increasing the silicon slab thickness, as shown in Fig. 2. The convergence study is performed within PBE. As can be seen, 6 and 4 layers of silicon are enough to achieve 0.02 eV accuracy of the charge transfer barrier for PO and NO configurations, respectively. In NO configuration, the charge transfer barrier is negative, thus there is more charge transfer than that of PO structure. The resulting larger screening effect in NO configuration facilitate its convergence. Therefore, we use 6 and 4 layers of silicon in the slab model for the PO and NO configurations respectively through all of the calculations. Since the PBE finite size corrections are about 2%, we expect the corrections to the GW calculations to be of the same order.

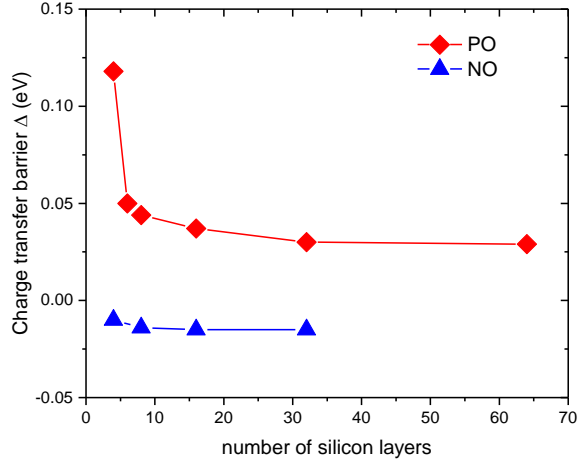

FIG. 2 The charge transfer barrier convergence as a function of the number of silicon layers. The red diamonds and blue triangles denote the PO and NO configurations, respectively.

## III. RESULTS AND DISCUSSIONS

TABLE I The charge transfer at Si(100):H-F4TCNQ interface, PO configuration, obtained by different methods. CDD stands for charge density difference method and DOSI stands for DOS integral.

| Method | Mulliken | Hirshfeld | Voronoi | Bader | CDD | DOSI | Chemical hardness |
|---|---|---|---|---|---|---|---|
| Charge ($e$) | 0.14 | 0.13 | 0.16 | 0.20 | 0.18 | 0.22 | 0.20 |

First, we compare the charge transfer calculated by the chemical hardness method with those of different charge analysis methods, namely the Mulliken[43], Hirshfeld[44], Voronoi[45] and Bader[46] charges, as shown in Table I. In addition, the charge transfer can be obtained by calculating the difference of the electron densities of the system before and after contact: $\Delta\rho = \rho_{AB} - \rho_A - \rho_B$. $\rho_{AB}$, $\rho_A$ and $\rho_B$ are the electron densities of the hybrid system, isolated A and B subsystems, respectively. The net charge transfer is $\delta n = \int \Delta\rho z dz$, where z is the spatial coordinate perpendicular to the interface. We denote this method as charge density difference (CDD) method. For the hybrid systems with physisoption, one can also calculate the charge transfer by integrating the density of states (DOS):

$$\delta n = \begin{cases} \int_{-\infty}^{VBM} f(E) \text{DOS}(E) dE, & \text{for electrons} \\ \int_{CBM}^{+\infty} [1 - f(E)] \text{DOS}(E) dE, & \text{for holes} \end{cases} \quad (6)$$

where *f* is the Fermi-Dirac distribution function. This method is denoted as DOSI, in Table I. In order to compare the different methods at the same level, we employ Siesta[47] code which is capable of employing all the methods discussed above. As a test, we calculate the charge transfer at Si(100):H-F4TCNQ interface with PO configuration. As can be seen from Table I, all methods give similar results. Note, that this is only true for physisorption case. In the chemical hardness method, The chemical potential and chemical hardness can be obtained by the EA and IP as $\mu = -(EA+IP)/2$, $\eta = IP-EA$. The IP and EA of molecules are calculated using the $\Delta$SCF method, which is also extended to solids.[48] For the Si(100):H-F4TCNQ interface, the capacitance term $1/C_{AB}$ in Eq. (3) is calculated to be 0.4 eV, which is much smaller compared to the band gaps of 2.0 eV and 4.0 eV for Si(100):H and F4TCNQ, respectively. Therefore, we neglect the capacitance term in Eq. (3) for the charge transfer in silicon-molecule systems. Since the capacitance term is neglected, we end up with an upper bound estimate of the charge transfer with the chemical hardness method. In table I and II, IP=5.17 eV and EA=4.05 eV are used for silicon.

TABLE II The charge transfer $\delta n$ per molecule at silicon-molecule interfaces. IP: ionization potential. EA: electron affinity. The IP and EA data of F2HCNQ and F4TCNQ in gas phase are from GW calculations in the present work. For other molecules, we adopt the experimental data from Ref. [49] and its references.

| Molecule | IP (eV) | EA (eV) | δn (e) | Chemical structure |
|---|---|---|---|---|
| F2HCNQ | 9.420 | 4.890 | 0.450 | 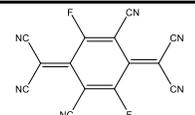 |
| F4TCNQ | 9.140 | 4.420 | 0.372 | 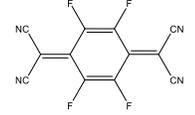 |
| TCNE | 11.770 | 3.160 | 0.293 | 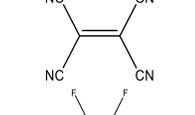 |
| F4benzoquinone | 10.700 | 2.450 | 0.210 | 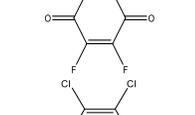 |
| Cl4benzoquinone | 9.740 | 2.754 | 0.202 | 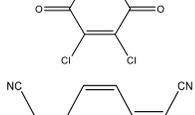 |
| TCNQ | 9.610 | 2.800 | 0.201 | 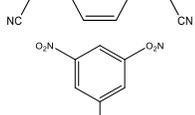 |
| dinitro-benzonitrile | 10.930 | 2.160 | 0.196 | 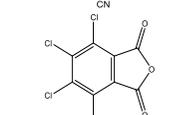 |
| Cl4-isobenzofurandione | 10.800 | 1.956 | 0.177 | 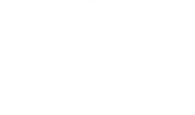 |

| Molecule | | | | |
|---|---|---|---|---|
| F4-benzenedicarbonitrile | 10.650 | 1.890 | 0.168 | 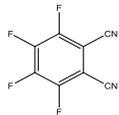 |
| PTCDA | 8.200 | 3.100 | 0.167 | 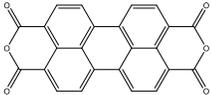 |
| maleic anhydride | 11.070 | 1.400 | 0.151 | 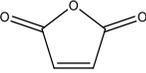 |
| Dichlone | 9.500 | 2.210 | 0.148 | 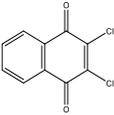 |
| Fumaronitrile | 11.150 | 1.249 | 0.144 | 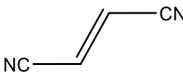 |
| Benzoquinone | 9.950 | 1.850 | 0.140 | 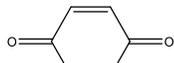 |
| nitro-benzonitrile | 10.200 | 1.687 | 0.138 | 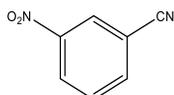 |
| Naphthalenedione | 9.400 | 1.800 | 0.114 | 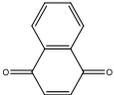 |
| phthalic anhydride | 10.100 | 1.245 | 0.107 | 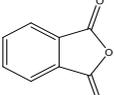 |
| C60 | 7.690 | 2.680 | 0.094 | 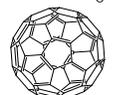 |
| mDCNB | 10.200 | 0.910 | 0.091 | 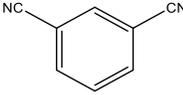 |
| Nitrobenzene | 9.860 | 1.000 | 0.082 | 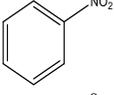 |
| Phthalimide | 9.780 | 1.015 | 0.080 | 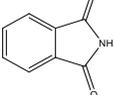 |
| NDCA | 8.980 | 1.260 | 0.058 | 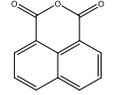 |
| Benzonitrile | 9.700 | 0.256 | 0.035 | 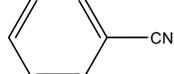 |

Table II summarizes the charge transfer between silicon and some well-known molecular acceptors calculated by Eq. (3). In the chemical hardness method, the driving force of the charge transfer is the chemical potential difference. Electrons transfer from higher chemical potential to the lower one. The resisting force of the charge transfer is the chemical hardness. It is hard to withdraw electrons from chemically hard materials. For example, the charge transferred from F2HCNQ (with chemical hardness of 4.53eV) is much larger than that of benzonitrile (with chemical hardness of 9.44 eV). Among all acceptors, F2HCNQ and F4TCNQ stand out to hold more potential as surface dopants. More accurate results refer to first-principles calculations. In the next step, we perform first-principles DFT and GW calculations on the top molecule candidates, namely F4TCNQ and F2HCNQ. TCNQ is also included for comparison.

TABLE III The band gap $E_g$ and charge transfer barrier $\Delta$ of TCNQ, F4TCNQ and F2HCNQ on Si(100):H surface with PO and NO configurations. d is the smallest distance between the molecule and Si(100):H surface. $\delta n$ is the charge transfer in units of cm$^{-2}$ and $e$ per molecule. The gas phase IP and EA of the molecules are shown for comparison. Both the PBE and GW results are presented. We do no show the charge transfer for PO configuration as it is too small due to the large barriers.

| molecule | method | Gas phase | | | On Si(100):H/PO | | | On Si(100):H/NO | | | | |
|---|---|---|---|---|---|---|---|---|---|---|---|---|
| | | IP (eV) | EA (eV) | $E_g$ (eV) | $d_{PO}$ (Å) | $E_g$ (eV) | $\Delta$ (eV) | $d_{NO}$ (Å) | $E_g$ (eV) | $\Delta$ (eV) | $\delta n$ (cm$^{-2}$) | ($e$) |
| TCNQ | PBE | 7.02 | 5.60 | 1.42 | 2.32 | 1.41 | 0.05 | 0.84 | 1.11 | -0.01 | 4.0×10$^{13}$ | 0.24 |
| | GW | 8.86 | 3.99 | 4.87 | | 2.17 | 0.77 | | 1.54 | 0.34 | 2.0×10$^{12}$ | 0.01 |
| F4TCNQ | PBE | 7.31 | 5.98 | 1.33 | 2.07 | 1.31 | 0.04 | 0.90 | 1.19 | -0.01 | 5.7×10$^{13}$ | 0.34 |
| | GW | 9.14 | 4.42 | 4.72 | | 1.46 | 0.67 | | 1.37 | 0.17 | 1.3×10$^{13}$ | 0.08 |
| F2HCNQ | PBE | 7.67 | 6.36 | 1.31 | 1.27 | 1.23 | 0.03 | 1.32 | 0.90 | -0.06 | 5.4×10$^{13}$ | 0.32 |
| | GW | 9.42 | 4.89 | 4.53 | | 1.37 | 0.64 | | 1.13 | 0.12 | 2.5×10$^{13}$ | 0.15 |

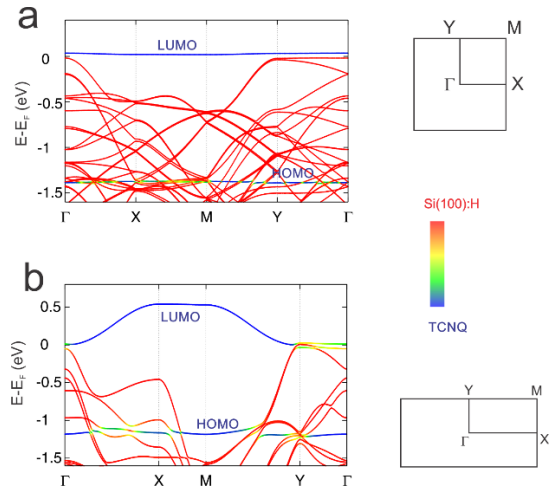

FIG. 3 Bands decomposition of Si(100):H-TCNQ in Wannier function basis as for (a) PO and (b) NO configurations. The color bar shows the relative weights of Si(100):H (red) and TCNQ (blue). On the right panel of each figure, the high-symmetry k points of the corresponding first Brillouin zone is displayed.

The main results of the DFT and GW calculations are summarized in Table III. In what follows, we take TCNQ as an example to present the GW calculations. The hydrogen-terminated silicon (100) surface with 1×1 reconstruction[50] Si(100):H is used in all the calculations. Figs. 1(a) and 1(b) show the structure of TCNQ on Si(100):H surface as in the parallel orientation (PO) which is found to be the most stable configuration of TCNQ molecule or SAM on silicon surface.[25,27] The smallest distance $d_{PO}$ between TCNQ and the Si(100):H surface is 2.32 Å and no chemical bonds form at the interface. We construct the bands decomposition, as shown in Fig. 3(a), of the hybrid structure in terms of maximally localized Wannier functions (MLWFs)[51,52]. The Wannier function centers are shown as the small red balls in Fig. 4(a). As the Wannier functions near the interface are well separated, the band structures of each part of the hybrid system is clearly identified. The HOMO and LUMO levels of TCNQ can be classified as the blue curves in Fig. 3(a), from which the PBE HOMO-LUMO gap of TCNQ in the PO configuration is extracted to be 1.41 eV which is nearly the same as 1.42 eV of the gas phase value, see Table III. Table IV shows the comparison of PBE and GW band energies at Γ point. The GW band gap of TCNQ reduces from 4.87 eV to 2.17 eV when approaching the silicon surface, indicating a large polarization induced screening due to interfacial charge transfer. The GW corrections shift the LUMO of TCNQ and the VBM of silicon up and down, respectively, increasing the charge transfer barrier from 0.05 eV to 0.77 eV which is too high for effective charge transfer at room temperature, since the thermal excitation energy is around 0.025 eV.

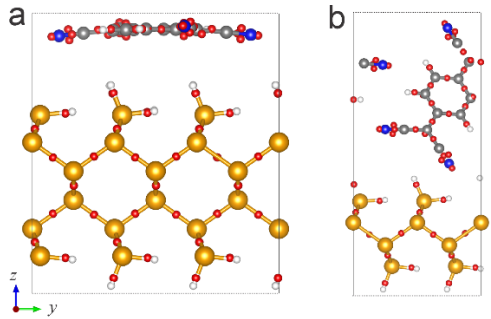

FIG. 4 Wannier function centers of Si(100):H-TCNQ, as denoted by the small red balls, in the (a) PO and (b) NO configurations.

TABLE IV Band energies (eV) of Si(100):H-TCNQ in PO configuration.

| Band index | PBE | GW | comment |
| --- | --- | --- | --- |
| 143 | -1.70534 | -2.335440 | HOMO of TCNQ |
| 144 | -1.69934 | -2.328667 | |
| 145 | -1.66713 | -2.327085 | |
| 146 | -1.66645 | -2.317321 | |
| 147 | -1.52264 | -2.220151 | |
| 148 | -1.50686 | -2.359260 | |
| 149 | -1.50343 | -2.327127 | |
| 150 | -1.49827 | -2.369440 | |
| 151 | -1.49078 | -2.301096 | |
| 152 | -1.35455 | -2.001875 | |
| 153 | -1.35221 | -1.997457 | |
| 154 | -1.11242 | -1.788307 | |
| 155 | -1.08982 | -1.760710 | |
| 156 | -0.92551 | -1.547314 | |
| 157 | -0.78377 | -1.405769 | |
| 158 | -0.78260 | -1.404522 | |
| 159 | -0.76992 | -1.457218 | |
| 160 | -0.52009 | -1.162392 | |
| 161 | -0.51192 | -1.153865 | |
| 162 | -0.34554 | -0.947526 | VBM of Si(100):H |
| 163 | -0.29474 | -0.168144 | LUMO of TCNQ |
| 164 | 1.18901 | 1.391024 | |

Putting molecules in a solvent, for example in molecular crystal, gap renormalizes due to electronic polarization effects.[53-55] Within the non-local vdW-DF-cx functional, the relaxed lattice parameters of TCNQ crystal are: a=8.86 Å, b=7.19 Å, and c=15.78 Å, which agrees well with experiment[56]. Figs. 5 and 6 show the TCNQ crystal structure and the corresponding HOMO and LUMO energies. Bulk gaps are aligned to the middle of the gas-phase gap. The bulk band gap reduces to 2.92 eV compared to 4.87 eV of the gas phase. The HOMO and LUMO levels of the crystal phase are shifted up and down, respectively. The resulting larger electron affinity is favored for hole injection. To this end, we construct the hybrid structure with a compact stack of TCNQ including strong inter-molecular interactions, as shown in Figs. 1(c) and 1(d). The distance $d_{NO}$ between the nitrogen atoms and silicon surface is as small as 0.84 Å. However, the N atoms are right on top of the vacuum inside the four H atoms. The Wannier functions are also well separated. Still there is no chemical bond forming at the interface, as shown in Fig. 4(b). We denote this configuration as nitrogen-terminated orientation (NO). The idea is to increase the polarization effect in the TCNQ environment and lower the LUMO level. We note that the goal can also be achieved by stacking up more TCNQ molecules in PO configuration on top of the silicon surface. However, it will increase the computational burden explosively. Moreover, the TCNQ in NO configuration can also be considered as one lamella as in its crystal phase, see Fig. 5. So the result of this configuration can serve as a good reference to TCNQ thin films. Fig. 3(b) shows the bands decomposition in terms of MLWFs. The dispersion of TCNQ bands along ΓX direction indicates strong inter-molecular coupling which can be seen from the face to face wave function overlap between the TCNQ molecules, as shown in Fig. 7. The band gap of TCNQ in NO configuration is corrected from 1.11 eV of PBE to 1.54 eV by GW calculations. The LUMO of TCNQ touch the VBM of silicon at Y point within PBE, as shown in Fig. 3(b). The charge transfer barrier is increased to 0.34 eV within GW, see Table III.

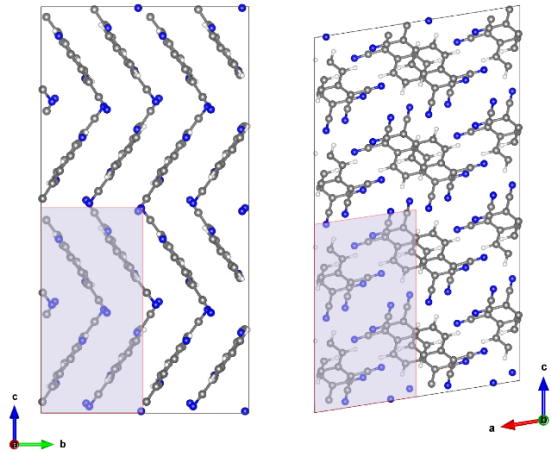

FIG. 5 Crystal structure of TCNQ. The shaded areas show the unit cell.

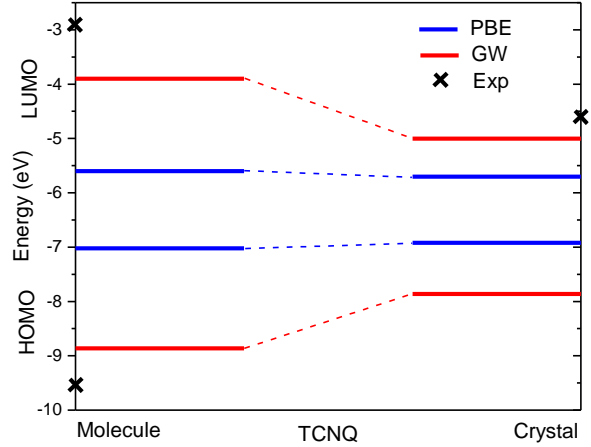

FIG. 6 The HOMO and LUMO energies of TCNQ in gas and crystal phases. The experimental data is taken from Ref[57-59].

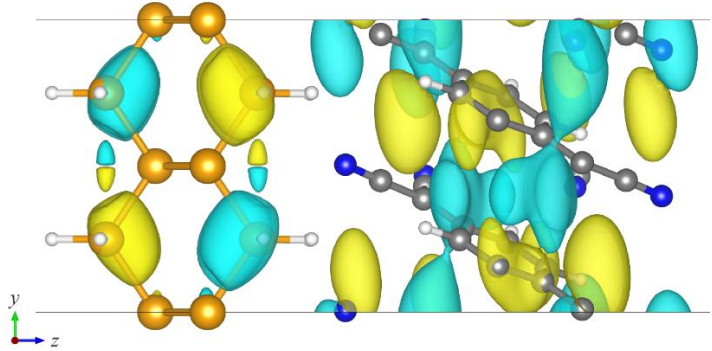

FIG. 7 The wave function of Si(100):H-TCNQ in NO configuration at Γ point for the energy corresponding to the LUMO of TCNQ. Different colors of the isosurfaces denote opposite signs of the wave function.

The carrier density can be calculated by Eq. (6). The GW DOS is approximated as a rigid shift of the DFT DOS by the corrected charge transfer barrier, since the GW corrections on the bands near the frontier orbitals which dominate the charge transfer are quite similar, see Table IV. The Fermi level in the GW DOS is set to be at the point where the amount of holes in silicon equals to that of electrons in TCNQ. Figs. 8(a) and 8(b) display the local density of states (LDOS) by PBE and GW, respectively. The GW correction removes the overlap of the LDOS in PBE calculations. The hole concentrations of silicon are calculated to be $4.0\times10^{13}$ cm$^{-2}$ and $2.0\times10^{12}$ cm$^{-2}$ by PBE and GW, respectively.

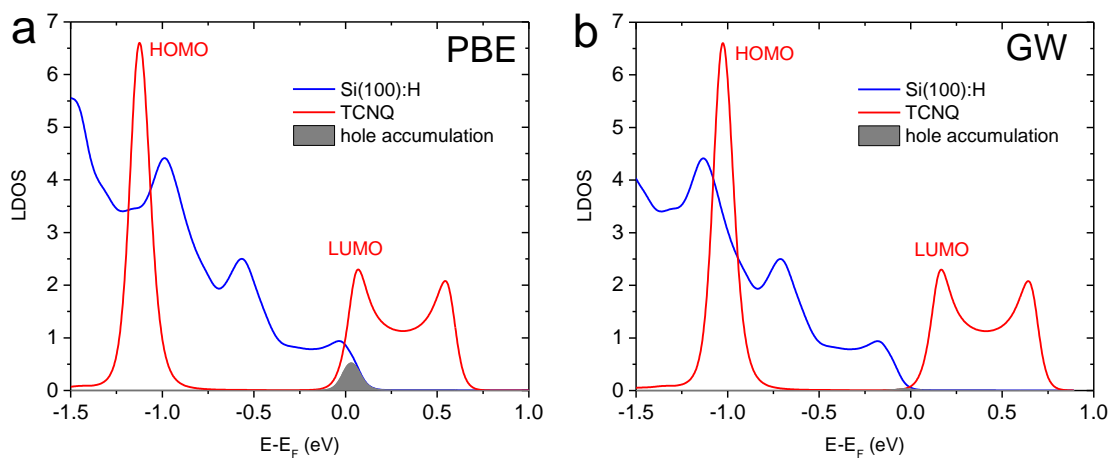

FIG. 8 The local density of states (LDOS) of Si(100):H-TCNQ in NO configuration for (a) PBE and (b) GW calculations, respectively. The GW LDOS is obtained by rigidly shifting the PBE LDOS with the corrected band offset. The shaded areas indicate hole accumulation.

The calculation details of F4TCNQ and F2HCNQ are quite similar to that of TCNQ, since all the three molecules share a similar structure, see Table II. The EA increases in the order of TCNQ < F4TCNQ < F2HCNQ, see Table III, consistent with the electron-withdrawing ability H < F < CN. The charge transfer barriers decrease as the EA increases, both for PO and NO configurations. Even for the strongest acceptor, F2HCNQ, the barrier in PO configuration is as large as 0.64 eV, still too high for effective charge transfer. For the NO configuration, both F4TCNQ and F2HCNQ can dope silicon effectively with hole concentrations of $1.3 \times 10^{13}$ cm$^{-2}$ and $2.5 \times 10^{13}$ cm$^{-2}$, respectively. So the F2HCNQ or F4TCNQ SAMs with lower coverage will not be able to dope silicon effectively. Instead, one would expect to use higher surface coverage or even thin films to achieve good performance in the silicon-organic devices.

## IV. CONCLUSION

In summary, we investigated charge transfer at silicon-organic interfaces by a combinatorial algorithm merging a thermodynamic method and first-principles DFT and GW calculations. A diversity of molecular acceptors are screened by using the simple and approximate thermodynamic method. The filtered candidates are further studied by more accurate first-principles GW calculations, capturing the dynamic polarization effects which are significant for HIO systems. The GW method can give more accurate band gap and band alignment than normal DFT calculations. We find that both F4TCNQ and F2HCNQ SAMs can efficiently dope silicon, achieving hole concentrations as high as $10^{13}$ cm$^{-2}$. The combinatorial algorithm used in the present work is quite general and could be applied to other HIO systems and the results provide important guidance for high performance HIO device design.

## ACKNOWLEDGEMENTS

The authors wish to acknowledge SOE HPC cluster of Rutgers and XSEDE, which is supported by National Science Foundation grant number ACI-1053575, for providing the computation resources. This work is supported by National Science Foundation grant number 1403089 (K.E) and grant number 1400246. (M.Z). We would also like to thank Dr. Junxi Duan for useful discussions.